# EVALUATION OF THE THERMAL AND HYDRAULIC PERFORMANCES OF A VERY THIN SINTERED COPPER FLAT HEAT PIPE FOR 3D MICROSYSTEM PACKAGES

*Slavka Tzanova[1], Lora Kamenova[2], Yvan Avenas[2], Christian Schaeffer[2]*

[1]Technical University of Sofia, [2]INP Grenoble

## ABSTRACT

The reported research work presents numerical studies validated by experimental results of a flat micro heat pipe with sintered copper wick structure. The objectives of this research were to produce and demonstrate the efficiency of the passive cooling technology (heat pipe) integrated in a very thin electronic substrate that is a part of a multifunctional 3-D electronic package. The enhanced technology is dedicated to the thermal management of high dissipative microsystems having heat densities of more than 10W/cm2. Future applications are envisaged in the avionics sector.

A 2D numerical hydraulic model has been developed to investigate the performance of a very thin flat micro heat pipe with sintered copper wick structure, using water as a refrigerant. Finite difference method has been used to develop the model. The model has been used to determine the mass transfer and fluid flow in order to evaluate the limits of heat transport capacity as functions of the dimensions of the wick and the vapour space and for various copper spheres radii.

## 1. INTRODUCTION

Stacking electronic substrates in 3D packaging allows creating compact, lightweight and multifunctional electronic modules. This conception answers the main requirement of miniaturization of the electronics market and plays a key role in the microelectronics nowadays. Every configuration has specific thermal constraints and requires the implementation of adapted cooling techniques.

The reported research work presents numerical studies validated by experimental results of a flat micro heat pipe with sintered copper wick structure. The objectives of this research were to produce and demonstrate the efficiency of the passive cooling technology (heat pipe) integrated in a very thin electronic substrate that is a part of a multifunctional 3-D electronic package [1].

## 2. METHODOLOGY

The importance of the electronics cooling in the avionics sector continues to remain substantial due to ever-improving microelectronics technologies and the consequent elevated power consumption density.

The reported research work involves the investigation of thermal solutions for a double sided substrate, part of a multifunctional 3-D electronic module (Figure 1). The objective is to integrate heat pipes into the double-sided electronic slices in order to satisfy the requirements for the power dissipation. Heat pipes have proven their efficiency for many applications where high heat fluxes suppress the possibility of applying conventional cooling systems [2]. Sintered copper spheres have been used for the capillary wick structure and pure water as a working fluid.

Numerical investigations have been performed under steady-state conditions in order to optimize a first fabricated in the laboratory copper flat heat pipe. In detail, this paper aims to provide 2D hydrodynamic model [3], [4] for simulating the performance of the heat pipe system. Mathematical formulation of the equations that govern the physical laws inside the heat pipe is also presented. The numerical solution of the 2D model is further validated by experiments and has shown better promise over the one-dimensional models.

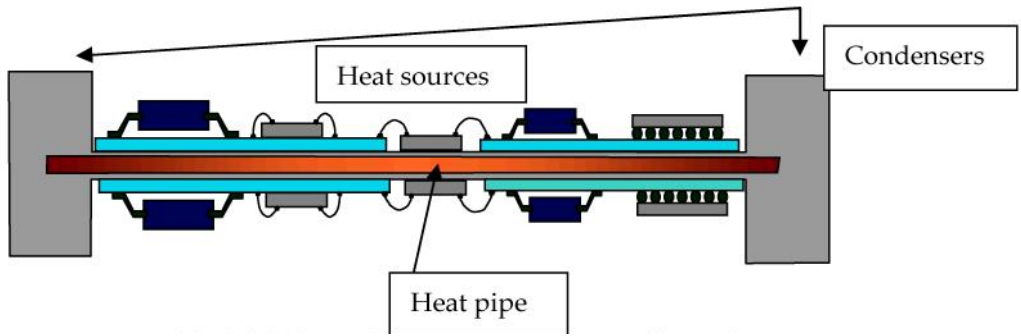

**Figure 1** Double sided substrate with integrated heat pipe

### 3. HEAT PIPE DESCRIPTION

The schematic diagram of the thin flat heat pipe is illustrated in Fig. 2. It consists of three basic sections - evaporator, condenser and adiabatic. Applying heat to the evaporator surface causes the liquid to vaporize. Subsequently, the pressure build-up entrains the vapour to move through the adiabatic section into the condenser, where it condenses. The liquid is then driven back to the evaporator by the capillary effect of the wick structure. Thanks to the isothermal characteristics of the vapour, there is a very small temperature gradient between the hot and the cold sections.

In the presented model, the hot and the cold sources are applied as shown on Fig. 2. The total length of the heat pipe is 44 mm; its width – 30 mm and height – 2.6 mm.

The wick structure within the heat pipe provides the capillary pressure needed for the return of the liquid from the condenser to the evaporator. To achieve maximum heat transport through the heat pipe, the geometry of the wick must be optimized. This requires trade-offs between different considerations. The maximum capillary pressure generated by the wick, which provides the liquid transport from the condenser to the evaporator, increases with the decrease of the effective pore size.

Another feature of the wick, which must be optimized, is its thickness. The heat transport capability of the heat pipe is raised by increasing the wick thickness. The 3D packaging requires that the thickness of the vapour core and the wick does not exceed 1mm ($h_v + h_w$ = 1mm, Fig. 2). The internal dimensions of the heat pipe are presented in Table 1.

Other important properties of the wick are the compatibility with the working fluid and the wet ability. In order to improve the wetting between the copper powder and the water, the copper wick structure was oxidized.

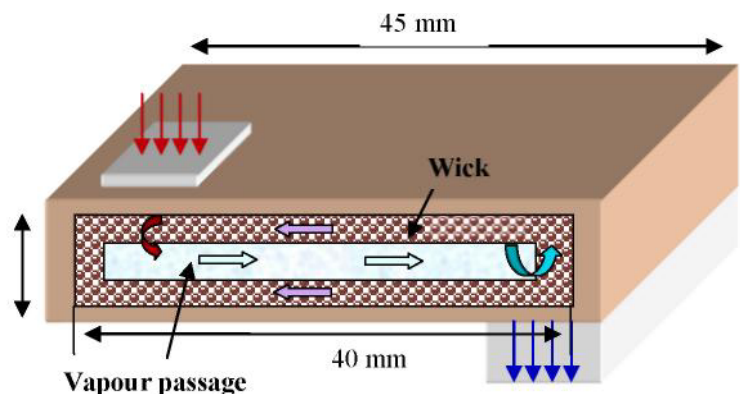

**Figure 2** A schematic of the section of the flat heat pipe

Table 1

|  | Material | Walls thickness | Wick | Vapour space |
|---|---|---|---|---|
| Heat pipe characteristics | Copper | 0.8 mm | 0.7 mm | 0.3 mm |

### 4. HYDRODYNAMIC 2D MODEL

The objective of the 2D hydraulic model presented here is to analyze the fluid flow and heat and mass transfer in the thin flat copper heat pipe and to predict the maximum heat transfer restricted by capillary limit. To estimate the limit, the vapor and the liquid pressure distributions within the device are needed. The basic physics of the fluid flow can be described by the equations of the mass and the momentum conservation for the vapor phase and the liquid phase. The conservation law means that for a control volume at steady state, the identity of the fluid within the control volume could change continuously, but the total amount present at any instant remains constant.

To solve the partial differential equations we apply finite difference numerical method that treats the pressure distribution at a finite number of grid points in the calculation domain [5]. The method includes a set of algebraic equations, solved repeatedly with updated coefficients, which leads to a solution that is sufficiently close to the correct solution. The finite difference technique is realized with Matlab.

The basic physics of fluid flow in the flat heat pipe can be described by the equations of mass and momentum conservation for the vapour phase and the liquid phase. The conservation law means that for a control volume at steady state, the identity of the fluid within the control volume changes continuously, but the total amount present at any instant remains constant. The mass rate balance is applied to a control volume and simplifications are made to do the analysis manageable. The control volume is modelled by making assumptions like steady-state flow, two-dimensionality (fluid velocity equals zero in z direction), boundary layer approximations and constant density laminar flow.

$\delta \dot{m}$ is the amount of the matter (mass flow rate) which enters the control volume ($\partial x \partial y h l$) during the evaporation process (Figure 3). The 2D model of the mass rate balance in the evaporator, adiabatic and condenser sections of the flat heat pipe is shown on Figure 4.

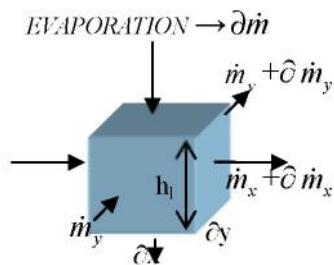

**Figure 3** Mass rate balance during evaporation

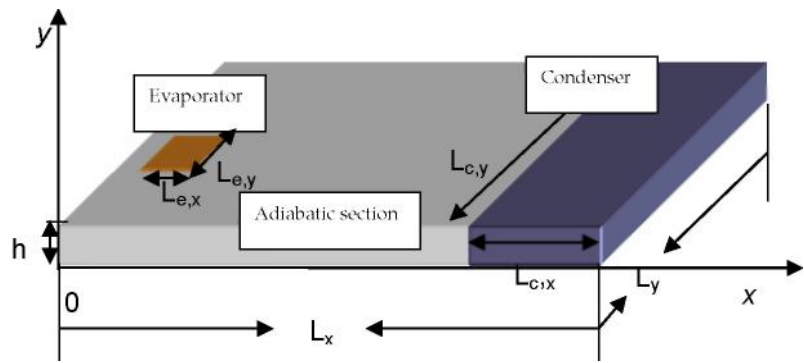

**Figure 4** Simulated heat pipe schematic

### 4.1. Capillary limit

During steady state operation of the heat pipe, the working fluid in the vapor phase flows continuously from the evaporator to the condenser, and it returns to the evaporator in the liquid phase thanks to the capillary structure. The heat pipe capillary limitation occurs when the capillary pressure difference is not enough to overcome the sum of the pressure drops in the liquid flow and the vapor flow. When exceeding the capillary limit, the effective thermal conductivity of the heat pipe reduces. The capillary limit formulation results from the Laplace-Young law (1).

$$(1) \quad \Delta P_{cap} = \frac{2\sigma cos\theta}{r_{eff}}$$

where $\sigma$ is the surface tension [N/m], $r_{eff}$ is the effective pore radius [m], $\theta$ - the wetting angle of the liquid vapor interface. Due to the good wetting between the oxidized copper spheres and the water, this angle should be very small. In our model we have chosen the wetting angle to be 10°.

### 4.2. Simulation results

The simulated model has the same dimensions and characteristics as the heat pipe used in the experimental setup. Capillary limit occurs when the maximum liquid pressure attains the minimum vapour pressure (we assume that the wetting angle *θ* in the condenser is zero).

$$(2) \quad \Delta P_{cap} = \Delta P_l + \Delta P_v = (P_{l,max} - P_{l,min}) + (P_{v,max} - P_{v,min})$$

Simulations of the performance of the model in terms of maximum heat power for working temperatures from 30°C to 80°C are presented on Figure 5. Very good prediction of the influence of the dimensions and the properties of the heat pipe is obtained.

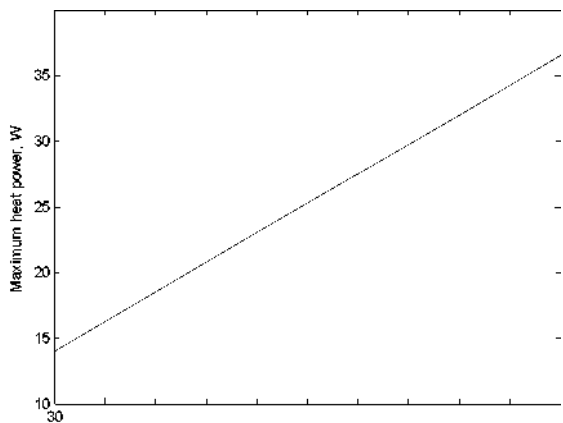

**Figure 5** Maximum heat power for various working temperatures

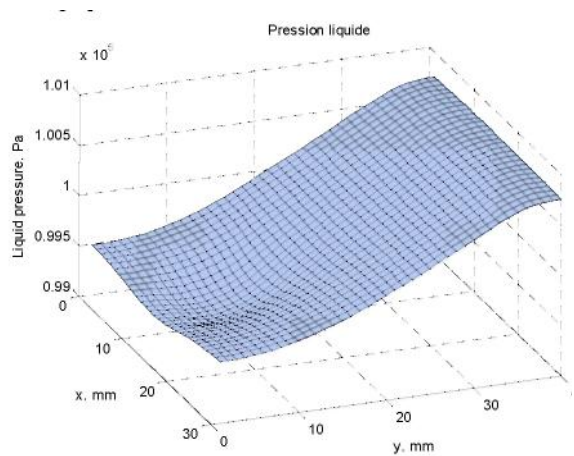

**Figure 6** Liquid pressure distribution in two-dimensional plan

### 4.3. Experimental results and discussions

The heat source consists of a silicone thermal resistance. The prototype is cooled by a cold plate realized by forced circulation of water at given temperature (Figure 7).
We applied different power inputs and we measured the temperature rise with an infrared camera and a thermocouple situated under the heat source, which gives more precise results (without taking account of the thermal contact resistance between the heat source and the heat pipe). The Figure 8 shows the effective thermal resistance values for several input powers. For each working temperature, the thermal resistance decreases until the value of the input power of 17-18W. From this value, the heat pipe was less effective because the capillary wick did not allow efficiently the return of the fluid to the evaporator. The capillary limit was attained.

The simulated results of the flat copper sintered heat pipe calculated with the 2D model presented here were compared to the results of the tested prototype, as well as with the simulations using the 1D model. The simulated results with the 2D model demonstrate better coherence to the experimental values. For example the simulated maximum heat power for working temperature of 60°C is 21W, while the experiment value was 17-18W. The difference between the simulations and the experiments is mainly due to the heat losses in the heat pipe envelope and because we give approximate values to some parameters, which are difficult to determine $(\theta, \varepsilon, h_l)$.

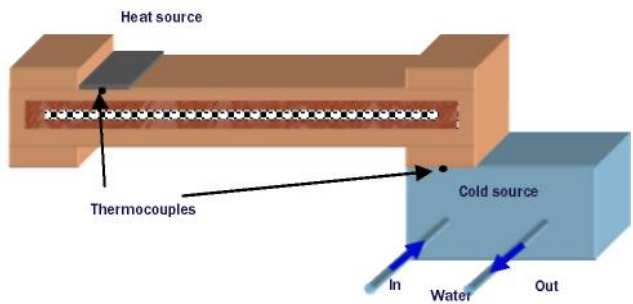

**Figure 7** Experimental set-up

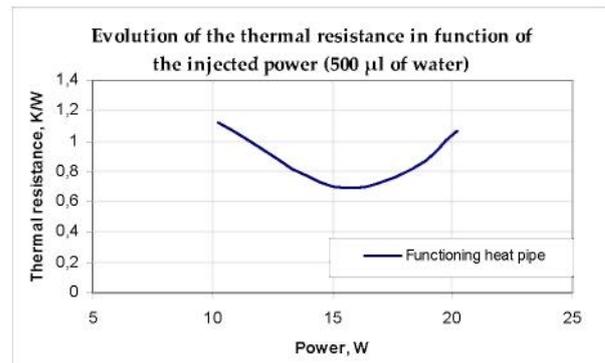

**Figure 8** Effective thermal resistance

## 5. CONCLUSION

In this paper we presented an effective cooling using heat pipes, integrated in the 3D module substrates. The developed simplified 2D hydraulic model of flat micro heat pipe with sintered copper wick structure has also been described. The results are presented in terms of liquid and vapor pressures and maximum heat power of the HP.

The simulated results with the 2D model demonstrate better coherence to the experimental values. The difference between the simulations and the experiments is mainly due to the heat losses in the heat pipe envelope and because we give approximate values to some parameters, which are difficult to determine The method can be used to predict the maximal heat power that the flat heat pipes could dissipate.